# Biomimetic Nanotechnology: A Powerful Means to address Global Challenges

I.C. Gebeshuber[1,2,*], B.Y. Majlis[1]

[1]Institute of Microengineering and Nanoelectronics
National University of Malaysia (UKM)
43600 Bangi, Selangor,
Malaysia

[2]Institute of Applied Physics & TU BIONIK Center of Excellence
Vienna University of Technology
1040 Vienna
Austria
*corresponding author ille.gebeshuber@mac.com

*Abstract:*
Biomimetic nanotechnology is a prominent research area at the meeting place of life sciences with engineering and physics: it is a continuously growing field that deals with knowledge transfer from biology to nanotechnology. Biomimetic nanotechnology is a field that has the potential to substantially support successful mastering of major global challenges.
The Millennium Project was commissioned by the United Nations Secretary-General in 2002 to develop a concrete action plan for the world to reverse the grinding poverty, hunger and disease affecting billions of people. It states 15 Global Challenges: sustainable development, water, population and resources, democratization, long-term perspectives, information technology, the rich-poor gap, health, capacity to decide, peace and conflict, status of women, transnational crime, energy, science and technology and global ethics. The possible contributions to master these challenges with the help of biomimetic nanotechnology will be discussed in detail.

## [1] Introduction

*1.1 The Millennium Project and the 'State of the Future' reports*
   The Millennium Project [1] was formed by the Futures Group International, the American Council for the United Nations University, the Smithsonian Institution and the United Nations University. The Millennium Project publishes the annual State of the Future report, now in its 12th edition (2008) and containing almost 100 print pages and around 6000 CD pages of data and analysis [2]. The 'State of the Future' report deals in detail with 15 major global challenges for humanity.

*"After 12 years of the Millennium Project's global futures research, it is increasingly clear that the world has the resources to address our common challenges. Coherence and direction are lacking. Ours is the first generation with the means for many to know the world as a whole, identify global improvement systems, and seek to improve such systems. We are the first people to act via Internet with like-minded individuals around the world. We have the ability to connect the right ideas to resources and people to help address our global and local challenges. This is a unique time in history. Mobile phones, the Internet, international trade, language translation, and jet planes are giving birth to an interdependent humanity that can create and implement global strategies to improve the prospects for humanity."*

[2] (from the executive summary, State of the Future 2008, [3])

*1.2 The 15 Global Challenges*

**Global Challenge 1: Sustainable Development**
- Heating and cooling without fossil fuels
- Microbial fuel cells
- Chemical heat production inspired by heat producing plants

**Global Challenge 2: Water**
- Deasalination with aquaporins
- Desalination via transpiration and filtering through membranes (inspired by mangroves)
- Water collection inspired by desert plants

**Global Challenge 6: Information Technology**
- New bioinspired optimisation processes (replacing slow trial and error models)
- Bioinspired methods of decision making (swarm intelligence)
- Management of over-information (technoinspiration)

**Global Challenge 8: Health**
- Nanomedical approaches inspired by immune system
- Biomimetic MEMS sensors
- Targeted drug delivery

**Global Challenge 13: Energy**
- Bioinspired wave and tide power systems
- Artificial photosynthesis
- Bioinspired hydrogen chemistry for fuel cells

**Global Challenge 14: Science and Technology**
- Overcoming the gap between the world of ideas, iinvestigators, inventors and innovators
- New dynamic ways of publishing
- New ways to access human knowledge (knowledge tree)

*Figure 1. Possible biomimetic contributions to help address selected global challenges as treated in Gebeshuber, Gruber and Drack, 2009 [4]*

Since 1997, the Millennium Project of the United Nations University has defined and tracked 15 global challenges in their annual State-of-the-Future reports. The challenges are updated each year. The 15 Global Challenges in the 2008 State-of-the-Future report [2] are:

**Sustainable Development** - How can sustainable development be achieved for all? **Water** - How can everyone have sufficient clean water without conflict? **Population and Resources** - How can population growth and resources be brought into balance? **Democratization** - How can genuine democracy emerge from authoritarian regimes? **Long-Term Perspectives** - How can policymaking be made more sensitive to global long-term perspectives? **Information Technology** - How can the global convergence of information and communications technologies work for everyone? **Rich-Poor Gap** - How can ethical market economies be encouraged to help reduce the gap between rich and poor? **Health** - How can the threat of new and re-emerging diseases and immune microorganisms be reduced? **Capacity to Decide** - How can the capacity to decide be improved as the nature of work and institutions change? **Peace and Conflict** - How can shared values and new security strategies reduce ethnic conflicts, terrorism, and the use of weapons of mass destruction? **Status of Women** - How can the changing status of women help improve the human condition? **Transnational Crime** - How can transnational organized crime networks be stopped from becoming more powerful and sophisticated global enterprises? **Energy** - How can growing energy demands be met safely and efficiently? **Science and Technology** - How can scientific and technological breakthroughs be accelerated to improve the human condition? **Global Ethics** - How can ethical considerations become more routinely incorporated into global decisions?

In a previous publication, Gebeshuber, Gruber and Drack investigated the possible contributions of biomimetics to mastering selected global challenges (Figure 1) that were identified to benefit greatly from this interdisciplinary approach [4].

Biomimetic contributions to expanding the potential for scientific and technological breakthroughs (Global Challenge 14, Figure 1) for example comprise overcoming the gaps between the world of ideas, inventors, innovators and investors (Figure 2) [4], new dynamic ways of publishing [5] and new ways to access human knowledge [5].

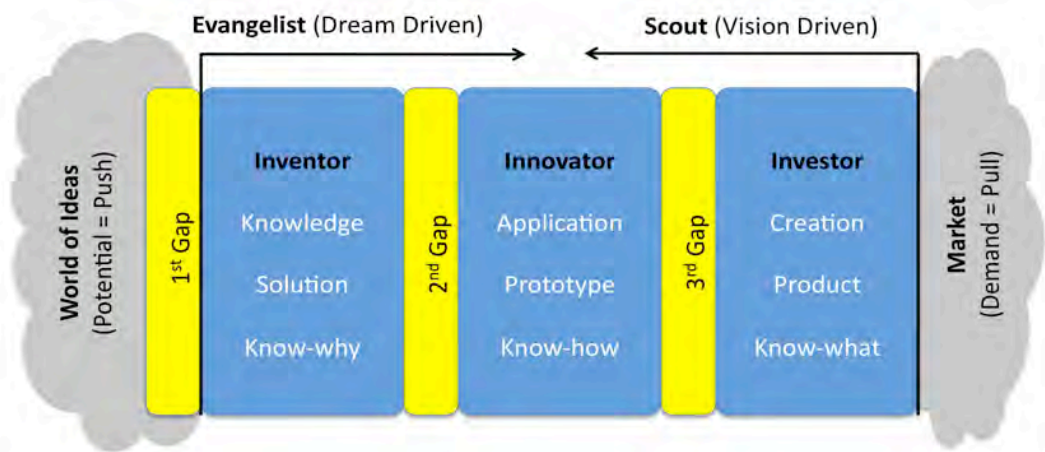

*Figure 3. The Three-Gaps-Theory proposed by Gebeshuber, Gruber and Drack in 2009 [4]. Image © 2009 Professional Engineering Publishing, UK. Image reproduced with permission.*

*Table 1: Life's principles as established by the Biomimicry Guild [7].*

| **Life's Principles - Design Lessons From Nature** |
|---|
| 1. Earth's operating conditions: |
|     1.1. Water-based |
|     1.2. Subject to limits and boundaries |
|     1.3. In a state of dynamic non-equilibrium |
| 2. Life creates conditions conducive to life |
|     2.1. Optimizes rather than maximizes |
|         2.1.1. Multi-functional design |
|         2.1.2. Fits form to function |
|         2.1.3. Recycles all materials |
|     2.2. Leverages interdependence |
|         2.2.1. Fosters cooperative relationships |
|         2.2.2. Self-organizing |
|     2.3. Benign manufacturing |
|         2.3.1. Life-friendly materials |
|         2.3.2. Water-based chemistry |
|         2.3.3. Self-assembly |
| 3. Life adapts and evolves |
|     3.1. Locally attuned and responsive |
|         3.1.1. Resourceful and opportunistic |
|             3.1.1.1. Shape rather than material |
|             3.1.1.2. Simple, common building blocks |
|             3.1.1.3. Free energy |
|         3.1.2. Feedback loops |
|             3.1.2.1. Antennae, signal, response |
|             3.1.2.2. Learns and imitates |
|     3.2. Integrates cyclic processes |
|         3.2.1. Feedback loops |
|         3.2.2. Cross-pollination and mutation |
|     3.3. Resilient |
|         3.3.1. Diverse |
|         3.3.2. Decentralized and distributed |
|         3.3.3. Redundant |

Here, in this publication, possible contributions of biomimetic nanotechnology to the remaining nine challenges are outlined. This is a complicated task, since besides the potential of biomimetic nanotechnology to serve society, further issues on the economic, sociological or political level have to be considered when addressing these nice challenges. Challenges No. 3 (Population and Resources), 4 (Democratization), 5 (Long-Term Perspectives), 7 (Rich-Poor Gap), 9 (Capacity to Decide), 10 (Peace and Conflict), 11 (Status of Women), 12 (Transnational Crime) and 15 (Global Ethics) fall into this category. In some of these challenges, biomimetic nanotechnology can contribute only very little at first sight. The biomimetic approaches suggested for these challenges can naturally not be as strong in terms of arguments, examples and possible technologies as the ones introduced in Gebeshuber, Gruber and Drack 2009, but they might prove inspiring and lead our minds on novel paths. And exactly this is the objective of this publication: to use a proved innovation method for the development of new ideas and approaches concerning major challenges of our world. The application of biomimetic nanotechnology that follows 'Life's principles' as introduced by the Biomimicry Guild [6] might help us to approach these problems in a sustainable way (Table 1).

As Gebeshuber, Gruber and Drack stated in 2009, the view expressed in the 15 Global Challenges is anthropocentric and a systems view of the biosphere should be taken to target ecological and sustainable development. This should also be reflected in the global challenges.

Biomimetic nanotechnology that follows 'Life's principles' (Table 1) as defined by the Biomimicry Guild from Helena, MT, USA [7] supposedly would yield sustainable products and processes. Sustainable biomimetic nanotechnology would thereby build from the bottom up, embrace diversity, self-assemble, adapt and evolve, optimize rather than maximize, use life-friendly materials and processes, use free energy, engage in symbiotic relationships, cross-pollinate and enhance the biosphere. These 'Life's principles' can serve as input for novel design approaches [6].

## 2. Possible contribution of biomimetic nanotechnology to the global challenges 3, 4, 5, 7, 9, 10, 11, 12 and 15

*2.1 Global Challenge 3: Population and Resources*

*Issues:* The gap in living standards between the rich and poor promises to become more extreme and divisive. Economic growth brings both promising and threatening consequences. Work, unemployment, leisure, and underemployment are changing.

*Opportunity:* There are same examples in nature that might provide inspiration on how to manage the rate of population growth. However, controlling the rate of population growth is a sensitive issue; and ethical as well as cultural issues have to be considered. Selected possible inspirations from nature that might subsequently be administered via nanomedicine methods such as nanoparticles or targeted drug delivery concern safer contraceptives based on studies of unusual food practices in people and animals [8] and selecting the sex of your offspring [9][10]. Control the sex of the offspring would prohibit the killing of newborn girls in places of the world with a dominant preference for male offspring. Apart from saving humans from being killed, the ability to predetermine the sex would also benefit chicken fledglings: currently, worldwide annually one billion male chicken fledglings are killed on the first day of their lives. They are not used in animal production since they would not lay eggs and would gain less weight than their female counterparts.

*2.2 Global Challenge 4: Democratization*

*Issue:* How can genuine democracy emerge from authoritarian regimes?

*Opportunity:* There are various examples for biomimetics contributions to transforming authoritarian regimes to democracies. Game theory models dealing with the evolution of human cooperation [11] and various processes in social insects such as bees [12] are two examples from nature relating to democracy. The political economy of bee swarming offers a fascinating study of collective action in biological systems [13]. Honeybee colonies possess decentralized decision-making, combining effectiveness with simplicity of communication and computation within a colony [14]. In 2004, Seeley and co-workers showed that groups of honeybees vote on hive locations via quorumsensing [15]. Honeybee inspired products and application ideas can be envisaged in control mechanisms for nanorobots and nanodevices that have to make decisions, as well as in government, community planning and business management. The bees' rules for decision-making - seek a diversity of options, encourage a free competition among ideas, and use an effective mechanism to narrow choices (range voting) - so impressed Seeley that he currently uses them at Cornell University as chairman of his department.

*2.3 Global Challenge 5: Long-Term Perspectives*

Issue: How can policy-making be made more sensitive to global long-term perspectives?

Opportunity: The fact that nature works solution oriented could be inspiration for bridging the gap between the short-term perspectives of policy-makers and the long-term perspectives of planers. In natural systems, solutions that work best for the whole system are favoured. Nature reacts to necessities for the whole, as opposed to maximum benefits for single individuals. Instead of multiple compromises to single individuals, the best possible solution for the whole system should be applied. Not the compromise is important, but long-term planning. Maximum benefit and not minimum hassle should be aspired. In this context it could be said that the sum of all compromises is death. Nature shows us that egocentric behaviour works only if it is the egocentricity of the whole system, not of the single individual. The global challenges have to be approached on a time frame that is longer than the terms of office of single decision-makers, longer than the lifetime of a single person. This leads to the question: What is better – one large solution or many small ones? Crucial for a global perspective in policy-making is finding the right frame for a decision, and the development of structures that enable decision-making. Biology can deliver role models for the negotiation between local and global decision-making, for example in the gradual dependence or independence of colony members [16][17]. The introduction of either distinct or integrated sensory and actuation micro- and nanosystems for global monitoring and action could be modelled according to relevant natural role models [18].

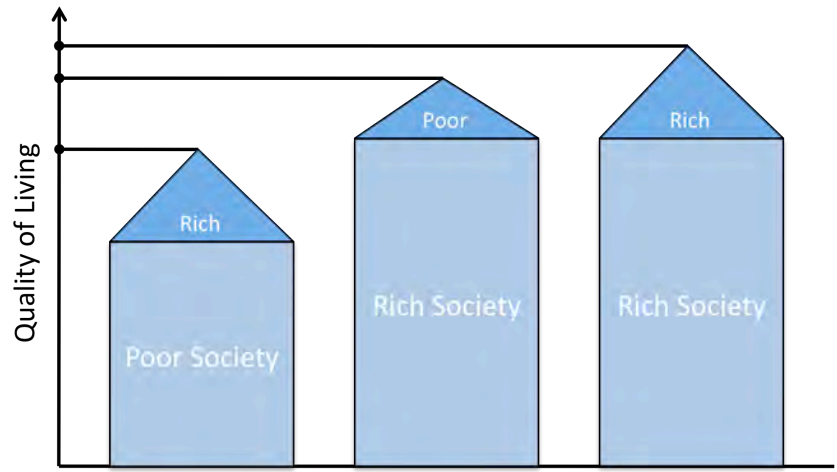

*Figure 4: Absolute richness of society (blocks) and relative richness of individuals (triangles).*

*2.4 Global Challenge 7: Rich-Poor Gap*

*Issue:* How can ethical market economies be encouraged to help reduce the gap between rich and poor?

*Opportunity:* Richness in nature is a different concept as richness for people. In the tropical forest, plants convert the energy provided by the sun and animals subsequently use this energy. We can learn from nature that richness means to take as much of the energy as needed. If we take more than we are entitled to, others get poor.

Richness in nature is relative, not absolute. Inspired by nature, the authors propose the concept of absolute and relative richness. Absolute richness is the richness of the whole society (block in Figure 4), and relative richness is just an add-on (pyramid that tops the block, Figure 4). Based on this concept, we propose an 'ethics of the rich' based on the fact that ensuring absolute richness of the whole society is beneficial, whereas voracious increase of the relative richness of single actors without contributing to society is not beneficial.

*2.5 Global Challenge 9: Capacity to Decide*

*Issue:* Diminishing capacity to decide (as issues become more global and complex under conditions of increasing uncertainty and risk).

*Opportunity:* Swarm intelligence can contribute to improve the capacity to decide as the nature of work and institutions changes. It is an example from nature that shows us how a population that is made up of simple agents interacting only locally with one another and with their environment can exhibit 'intelligent' global behaviour.

The agents follow very simple rules, and there is no centralized control structure dictating how individual agents should behave. Local, and to a certain degree random, interactions between such agents lead to the emergence of global behaviour. Natural examples of swarm intelligence include ant colonies, bird flocking, animal herding, bacterial growth and fish schooling. Mathematical models of social insects collective behaviour can be used in solving optimization problems [19]. This is of high relevance concerning emerging smart MEMS and NEMS. The behaviour of ants can be used to implement ACO (Ant Colony Optimization) algorithms to solve real-world problems including routing optimization, structure optimization, data mining and data clustering [19].

*2.6 Global Challenge 10: Peace and Conflict*

*Issues:* Increasing severity of religious, ethnic, and racial conflicts. Terrorism is increasingly destructive, proliferating, and difficult to prevent.

*Opportunity:* In many cases, nature is merciless. Drawing too straightforward conclusions from organisms and ecosystems for bioinspired ideas, applications and processes in terms of novel security strategies might clash with ethical values. One fact that nature does teach us is that freedom has its price, and high biodiversity is beneficial. Human alteration of the global environment has triggered the sixth major extinction event in the history of life and caused widespread changes in the global distribution of organisms [20][21].

Biomimetic micro- and nanosensors for crime prevention and detection are a hot topic of research [22] and the immune system can inspire novel defence systems: the biological immune system is an autonomic system for self-protection. The powerful information processing capabilities of this system include feature extraction, pattern recognition, learning and memory.

An autonomic self-adaptive defence system might be inspired by some immunological metaphors for information gathering, analysis, decision-making and launching threat and attack responses [23]. See also section on transnational crime.

*2.7 Global Challenge 11: Status of Women*
*Issue:* The status of women is changing.
*Opportunity:* Investigating the potential of biomimetics related to improving the status of women implies the assumption of a socially relevant difference between men and women, and cultures with 'male' or 'female' attributes, otherwise the question would be trivial. This is again a controversial theme, but in view of the authors there should be equilibrium between male and female cultures. Economic autonomy of women is important. In many male dominated societies men do not take responsibilities for children. Economic autonomy of women provides children with better access to resources. The main focus is equitable distribution of resources, especially for the 'structures' that depend on the women (children, parents). In nature, the access to resources is determined by demand and not by structures. Hierarchical distribution of resources in patriarchal systems is counterproductive. Growing importance of women in society will have three benefits: First, more direct access to resources, secondly, higher responsibility of men regarding women, including the joint raising of children, and thirdly, stronger networks amongst women. Still, in some cultures, women do not have the possibility to gather, to network, to publicly be seen and listened to.

*2.8 Global Challenge 12: Transnational Crime*
*Issue:* Transnational organized crime networks are becoming more powerful and sophisticated global enterprises.
*Opportunity:* Each structure and each society has a certain amount of energy and resources that are distributed internally. This distribution follows strictly defined rules and established structures. Crime networks live either on infiltration (and subsequent manipulation) or substitution of these structures. The solution of this problem is not to declare war on the whole system or parts of it – this can also be learned from history and the human body. Many of these networks are so closely tied to the host structure that they cannot be separated from it anymore without major harm for the organisms itself (e.g., cancer). If the critical mass of crime organisations is reached, whole nations become instable: they become 'failing nations' [24]. A basic interest concern of crime networks is to keep the host structures as an integral part of their system. The complexity of human structures allows for utilisation of powers of crime networks for strengthening the whole system and establishing a new order. In 2005, Alvaro Uribe, president of Colombia, imposed an amnesty over all the drug barons in his country, and even allowed them to apply for political functions. Opening a career path for little street criminals towards acceptance in society, enabling criminals to evolve to sustainers of the system prevents build-up of a critical mass of criminals, and therefore 'failing nations'. Mexico is currently on the path to becoming a 'failing nation' and effective strategies to prevent this are high in demand. The endosymbiotic theory concerns the origins of mitochondria and plastids (e.g. chloroplasts), which are organelles of eukaryotic cells. Mitochondria and plastids originated as separate prokaryotic organisms and were taken into the cell. In many instances of endosymbiosis either the endosymbiont or the host cannot survive without the other [25].

*2.9 Global Challenge 15: Global Ethics*
*Issue:* Need to encourage diversity and shared ethical values.
*Opportunity:* Ethics is stipulated moral. In highly ordered structures such as insect states it is sufficient to integrate 'ethics' in the hardware (instincts, natural behaviour). In more complex structures such as wolf packs, tribes, nations and economic societies ethics result from the interaction of cultures, from evolution of traditional patterns of behaviour (religious patterns) and from the necessity to live together. Our ethics is based on evolutionary basic principles. However, ethics and moral also depend on fashion, and can change with time, compare e.g. the view of white northern Americans against African Americans.

Nature teaches us that following a fixed ethical scaffold is very advantageous in living together. Instincts and consequently abiding to this ethics ensures survival of the whole. Adjusting ethics to individual interests is an infringement.

### 3. Conclusions

The high potential of biomimetic nanotechnology is a good reason for intensifying biomimetic research. The public image of biomimetics is positive, and if biomimetics increasingly aims at providing sustainable solutions, this image might not change.

Successful biomimeticians are inherently transdisciplinary thinkers. A common language for biologists and engineers, in which descriptions at different level of detail are more compatible, is needed [5].

Many of the currently envisaged solutions to the global challenges facing humanity are in paramount contradiction to the 'approach' of nature (e.g., the iron fertilization project for the oceans, LOHAFEX, [26]). A discussion about costs of stagnation *vs.* costs of growth (the survival of the fittest, permanent strife for resources) is needed. The question remains whether our current ethics are capable of grasping the complexity of the systems and of allowing solutions that are necessary in the long run. How can we state being ethical entities and concurrently adapt the destroying power of our weapon systems to the size and density of the population? On the other hand we cannot supply sustainable solutions and resources for the demand of our fellow beings, respectively. In light of these thoughts questions about the future of humanity might have to be reformulated. Considering real processes (environment-person-environment) the emerging field of biological physics shall pose the questions and human organizations shall answer them:
- How can we prevent a catastrophe for humankind (fast extinction *vs.* slow extinction)?
- How can resources be better distributed without the distribution system becoming the critical factor for survival (i.e., dependence on sensitive logistical systems)?
- How can the inevitable struggle between humans be moved to levels that do not constrain the basis of life?
- How can we prevent that people who have better access to resources use them quasi exclusively (i.e., fair distribution, social peace)?
- How can we accept that our systems and we ourselves are imperfect and that – even if we want to act ethically as individuals – we are forced to act unethically in society (i.e., limits of ethics)?

**Acknowledgements**

The Austrian Society for the Advancement of Plant Sciences funded part of this work via the Biomimetics Pilot Project 'BioScreen'. The authors thank Jennifer Bawitsch and Teresa Stemeseder for carefully reading the manuscript. The Biomimicry Institute, Montana, USA, is acknowledged for their great work, comprising extracting 'Life's principles' and for organizing biological literature by function in the open source project AskNature.com. In this way, they inspire technologies that create conditions conducive to life.